\documentclass[11pt,a4paper,oneside]{article}
\pdfoutput=1
\usepackage{jheppub}

\usepackage{verbatim}
\usepackage{mathrsfs}
\usepackage{appendix}
\usepackage{caption}
\usepackage{float}
\usepackage{subfig}
\usepackage[vcentermath]{youngtab}
\usepackage{alltt}
\usepackage{multirow}
\usepackage{arydshln}
\usepackage{lscape}
\usepackage{tabu}
\usepackage{amssymb,amsmath,amsfonts,mathrsfs,enumerate,wrapfig,latexsym,wasysym}

\usepackage{graphicx}
\usepackage{adjustbox}
\usepackage{xcolor}
\usepackage{slashed}    
\usepackage{url}
\usepackage{hyperref}
\hypersetup{
	colorlinks = true,
	linkcolor = blue,
	anchorcolor = blue,
	citecolor = blue,
	filecolor = blue,
	urlcolor = magenta,
	bookmarks=true,
	linktocpage=true
}
\usepackage{framed}
\usepackage[numbers,sort&compress]{natbib}
\usepackage{fancyvrb}
\allowdisplaybreaks
\usepackage{multirow}
\usepackage{framed}
\RequirePackage{flushend}
\usepackage{makecell}
\usepackage{color}
\usepackage{pifont}
\usepackage{cancel}
\usepackage{import}
\usepackage{subfiles}
\usepackage{filecontents}
\usepackage{rotating}
\usepackage{threeparttable,tablefootnote}\usepackage{mathtools}
\usepackage{pbox}
\usepackage{setspace}
\usepackage{lscape}
\usepackage{color, colortbl}
\usepackage{pdflscape}
\usepackage{afterpage}
\usepackage{fontawesome}

\usepackage{threeparttable}
\usepackage{slashed}
\usepackage[normalem]{ulem}

\definecolor{darkgreen}{cmyk}{1,0,1,0.4}

\definecolor{pink}{cmyk}{0.4,1,0.3,0}

\def\com2#1{\textcolor{red}{\it{#1}}}

\usepackage{xcolor}
\definecolor{gerua}{rgb}{0.74, 0.2, 0.64}
\definecolor{beguni}{rgb}{0.9, 0.17, 0.31}
\definecolor{darkgreen}{rgb}{0.07, 0.53, 0.03}
\definecolor{skyblue}{rgb}{0.0, 0.53, 0.74}

\graphicspath{{plots/}}

\newcommand{\xmark}{\ding{55}}
\definecolor{Gray}{gray}{0.9}

\preprint{IPPP/20/63}
\title{Classifying Standard Model Extensions Effectively with Precision Observables}

\author[1]{Supratim Das Bakshi,}
\author[1]{Joydeep Chakrabortty,}
\author[2]{Michael Spannowsky} \author{\\}

\affiliation[1]{Department of Physics, Indian Institute of Technology, Kanpur-208016, India}

\affiliation[2]{Institute for Particle Physics Phenomenology, Department of Physics, Durham University, Durham DH1 3LE, U.K.}

\emailAdd{sdbakshi, joydeep@iitk.ac.in, michael.spannowsky@durham.ac.uk}

\abstract{Effective theories are well established theoretical frameworks to describe the effect of energetically widely separated UV models on observables at lower energy scales. Due to the complexity of the effective theory when taking all the Standard Model symmetries and degrees of freedoms into account, tensioning the entire system in a completely agnostic way against experimental measurements results in constraints on the Wilson Coefficients of the effective operators that either bears little information or challenge intrinsic assumptions imposed on the effective field theory framework. In general, a specific high-scale extension of the Standard Model only induces a subset of all possible operators. Thus, by investigating which operators are induced by different classes of the Standard Model extensions and comparing to which precision observables they contribute, we show that it is possible to obtain an improved understanding of which UV model is realised in nature. We consider 15 UV models which are single scalar field extensions of the Standard Model and compute their dimension-6 operators after integrating out the heavy scalars up to 1-loop level. Only very few of these scenarios remain indistinguishable, while most of the models can be phenomenologically separated from one another. Most of these scenarios possess their own characteristic operator signature. Following the approach outlined here, a comparative analysis of a wide range of models will allow to assess at what level the effective field theory series can be truncated and which experimental measurements to prioritise.
					}

\begin{document}
	\maketitle
\section{Introduction}\label{sec:intro}

To explain the shortcomings of the Standard Model of particle physics, a plethora of extensions have been proposed that introduce new particles and interactions at high energies. Many of the anticipated new particles are outside of the kinematic reach of current and near-term high-energy experiments. To quantify their imprint on observables measurable at collider experiments, which have currently an energy reach up to $\mathcal{O}(1)$ TeV, effective field theories provide an established theoretical framework to tension data with hypothesised deformations of the Standard Model.

\noindent
In case all degrees of freedom of the UV model is too heavy to be excited on-shell, the most suitable framework, built exclusively on Standard Model particle content and symmetries, is the so-called Standard Model Effective Field Theory (SMEFT) \cite{Grzadkowski:2010es,BUCHMULLER1986621,Jenkins:2013zja,Jenkins:2013wua,Alonso:2013hga,Elias-Miro:2013gya,Elias-Miro:2013eta,Hartmann:2015oia,DasBakshi:2019vzr}. After integrating out the degrees of freedom that are introduced through the SM UV extension, the effective action can be expanded in terms of local operators of dimension-$n$, each suppressed by factors $\mathcal{O}(1/\Lambda^{n-4})$. At dimension-6, the SMEFT basis already consists of 2499 operators \cite{Grzadkowski:2010es,Alonso:2013hga}, most of them 4-fermion operators. Even if we assume that none of the operators act as a source for flavour violation, we are left with 59 operators. An entirely agnostic approach, i.e. a bottom-up approach that does not make any assumption about the UV theory, requires us to put all SMEFT operators on the same footing and tension them all simultaneously. Due to the complexity of this framework, over-constraining the entire system simultaneously by tensioning it against measurements is a highly challenging task \cite{Englert:2015hrx, Butter:2016cvz,Ellis:2018gqa,deBlas:2019rxi}. To date, observables have not been measured precise enough for all operators to be given an agnostic interpretation in terms of a weakly coupled UV theory \cite{Araz:2020zyh,deBlas:2019rxi}. Thus, to perform a global fit within the EFT framework, further assumptions have to be imposed on the system, e.g. that the Wilson coefficients are small enough not to violate perturbative unitarity within the energy ranges of the measurements \cite{Englert:2014cva,Englert:2017aqb}.

\noindent
Thus, for practical reasons, it seems prudent to augment EFT analyses by taking into account what we know about the hypothesised UV models. In general, a given UV model induces only a subset of the set of all SMEFT operators. Further, some operators might only be induced at loop level and should therefore be suppressed compared to tree-level induced operators. Some UV models might induce B and L violating processes, while others do not. Breaking degeneracies in how a UV model leaves an imprint on precision observables improves the constraints obtained from a global EFT fit tremendously \cite{Anisha:2020ggj}.

\noindent
In this paper, focusing on dimension-6 SMEFT operators throughout, we will attempt a first step towards providing a classification of UV scenarios using electroweak precision observables, Higgs phenomenology, perturbativity constraints and baryon ($B$) and lepton ($L$) number violating processes \cite{Grzadkowski:2010es}. 
Two important aspects for the categorisation are (i) the choice of suitable observables, and (ii) at what order one truncates the perturbative series in the calculation of the effective operators. 

\noindent
To showcase how a comprehensive consideration of phenomenological and theoretical constraints can result in an improved characterisation of possible UV theories from EFT operators alone, we apply this approach to 15 different single-particle extensions of the SM, covering a wider range of models from colored to uncolored scalars\footnote{While this appears to be a fairly constrained selection of BSM models, we note that extensions of the scalar sector are present in many BSM models \cite{Zhang:2016pja,Jiang:2018pbd,Dawson:2017vgm,Haisch:2020ahr,deBlas:2014mba,deBlas:2017xtg,Henning:2014wua,Ellis:2017jns,Deshpande:1977rw,Nie_1999,Branchina:2018qlf,Pilaftsis:1999qt,Babu:2009aq,Bambhaniya:2013yca,Bauer:2015knc,Bandyopadhyay:2016oif,Davidson_2010,Arnold:2013cva,Buchmuller:1986zs,Assad:2017iib,Chen:2008hh}. Further, we stress that if the envisioned BSM model extends the SM by multiple approximately degenerate degrees of freedom, at dimension-6, calculated up to 1-loop, the operators induced by integrating out each particle contribute linearly to the combined set of operators.}. After integrating out these heavy non-SM particles from each BSM Lagrangian, we compute the dimension-6 effective operators up to 1-loop level, including the mixing between light and heavy degrees of freedom \cite{Zhang:2016pja,Ellis:2016enq,Ellis:2017jns,Kramer:2019fwz,Ellis:2020ivx}. Some of these operators contribute to electroweak precision observables (EWPO) at leading (LO) or next-to-leading (NLO) order or to the Higgs signal strengths (HSS). A small number of operators do not contribute to either of these observables, which are then marked as so-called additional operators (AdOps) in our analysis. In turn, a few colored scalar extension BSM scenarios lead to $B, L$ violating (BLV) dimension-6 operators. Thus, we classify all of these effective operators into the following categories: (i) EWPO-LO, (ii) EWPO-NLO-I, (iii) EWPO-NLO-II, (iv) HSS, (v) AdOps, and (vi) BLV based on whether the UV theory induces the respective operators or not. As a result, this categorisation allows us to classify the 15 BSM models according to the operators they induce and how one can probe such a model phenomenologically.
Such categorisation can guide and inform future experimental measurements. This approach shows the importance of computing the operators of a given mass dimension as precisely as possible and highlights the need to enlarge the set of experimental measurements. The interplay between theory and experiment is of great benefit not only to identify new physics models but to provide guidelines for the prioritisation of experimental measurements. 

In the following, we aim to highlight the methodology and intricacies of BSM classifications using the SMEFT framework. First, in Sec.~\ref{sec:BSM-SMEFT}, we discuss how the heavy BSM fields are integrated out and we tabulate the emerged dimension-6 effective operators for each of the 15 UV models. Then, in Sec.~\ref{sec:obs-ops}, we show which operator contributes to which of the chosen precision observables. In Sec~\ref{sec:obs-bsm-class}, we briefly outline the methodology of our classification. We detail how operator degeneracies between models can be broken by incorporating an increasing number of observables successively. We note the cumulative effect of all the observables in the BSM classification and we present its final pattern in Sec.~\ref{sec:BSM-class-present-future}. We further argue the need of including effective operators beyond the 1-loop level, for which we need to improve the precision of theoretical computation. In Sec.~\ref{sec:conclusion} we offer a summary and conclusions.

\section{Integrating out heavy fields: BSM to SMEFT}\label{sec:BSM-SMEFT}

We have considered 15 example BSM scenarios which are single-particle extensions of the SM. All the adopted models contain different representations of scalar fields: (i) color-singlet $SU(2)_L$ singlet (real \& complex) and non-singlet \cite{Zhang:2016pja,Jiang:2018pbd,Dawson:2017vgm,Haisch:2020ahr,deBlas:2014mba,deBlas:2017xtg,Henning:2014wua,Ellis:2017jns,Deshpande:1977rw,Nie_1999,Branchina:2018qlf,Pilaftsis:1999qt,Babu:2009aq,Bambhaniya:2013yca}, and (ii) colored-non-singlets $SU(2)$  singlet and non-singlet complex scalars \cite{Bauer:2015knc,Bandyopadhyay:2016oif,Davidson_2010,Arnold:2013cva,Buchmuller:1986zs,Assad:2017iib,Chen:2008hh}, which are assumed to be sufficiently heavy to be integrated out leaving the SM as the low energy theory. The SM gauge quantum numbers of these fields are depicted in Table~\ref{tab:BSMs-op-obs}. We have integrated out these heavy non-SM fields from each of the BSM scenarios and computed exhaustive  sets of dimension-6 effective operators  up to 1-loop level using CoDEx \cite{Bakshi:2018ics}. We have tagged    the operators with T, HH, and HL based on whether they arise from heavy tree-propagator, all heavy loop-propagators \cite{Henning:2014wua,Drozd:2015rsp,Henning:2016lyp}, or heavy-light mixed loop-propagator \cite{Zhang:2016pja,Ellis:2016enq,Ellis:2017jns,Kramer:2019fwz,Ellis:2020ivx} respectively, while \xmark\; signifies the  absence of that particular operator for a given BSM scenario.

\begin{table}[h!]
	\caption{ Effective operators for 15 BSM scenarios in Warsaw basis: integrating out heavy-tree (T), heavy-loop (HH), and heavy-light-loop (HL) propagators.}
	\label{tab:BSMs-op-obs}
	\renewcommand*{\arraystretch}{3}
	\begin{adjustbox}{width=1.05\textwidth}
		\begin{tabular}{|p{30pt}|p{40pt}|c|c|c|c|c|c|c|c|c|c||c|c|c||c|c|c|c|c|c||c|c|c|c|c|c|c|c|c|c|c|c|c|c|c|c|c||c|c|c|c|c|c|c||c|c|c|c|}
			\hline
			& &1 &2 &3 &4 &5 &6 &7 &8 &9 &10 &11 &12 &13 &14 &15 &16 &17 &18 &19 &20 &21 &22 &23 &24 &25 &26 &27 &28 &29 &30 &31 &32 &33 &34 &35 &36 &37 &38 &39 &40 &41 &42 &43 & 44& 45& 46& 47\\
			\hline
			\small Heavy\newline BSM \newline fields& \large  \text{  } $\mathcal{G}_{3,2,1}$\newline  \newline & \color{blue} $Q_{HD}$ &\color{blue} $Q_{ll}$ & \color{blue}$ Q_{Hu} $ &\color{blue}$ Q_{Hd} $ &\color{blue} $ Q_{He} $ & \color{blue}$ Q_{Hq}^{(1)} $ &\color{blue}$ Q_{Hl}^{(1)} $ &\color{blue}$ Q_{Hl}^{(3)} $ &\color{blue}$ Q_{Hq}^{(3)} $ & \color{blue} $Q_{HWB}$ &\color{beguni} $Q_{H\square}$ &\color{beguni} $Q_{HB}$ &\color{beguni} $Q_{HW}$ &\color{darkgreen} $ Q_H$ &\color{darkgreen} $ Q_{G} $ &\color{darkgreen} $Q_{HG}$ &\color{darkgreen} $Q_{eH}$&\color{darkgreen} $Q_{uH}$ &\color{darkgreen} $Q_{dH}$ & \color{gerua} $Q_{qq}^{(1)}$ &\color{gerua} $Q_{qq}^{(3)}$ &\color{gerua} $Q_{uu}$ &\color{gerua} $Q_{dd}$ &\color{gerua} $Q_{ud}^{(1)}$ &\color{gerua}$ Q_{lq}^{(1)} $ & \color{gerua}$ Q_{ee} $&\color{gerua}$ Q_{eu} $ &\color{gerua}$ Q_{ed} $ &\color{gerua} $ Q_{le} $ &\color{gerua}$ Q_{lu} $ &\color{gerua}$ Q_{ld} $ &\color{gerua}$ Q_{qe} $ &\color{gerua}$ Q_{qu}^{(1)} $ &\color{gerua}$ Q_{qd}^{(1)} $  &\color{gerua} $ Q_{lq}^{(3)} $&\color{gerua} $Q_{W} $ & \color{skyblue} $Q_{ud}^{(8)}$ & \color{skyblue}$ Q_{qd}^{(8)} $ & \color{skyblue}$ Q_{qu}^{(8)} $ &\color{skyblue} $Q_{quqd}^{(1)}$ &\color{skyblue} $ Q_{lequ}^{(1)} $&\color{skyblue} $Q_{quqd}^{(8)}$ & \color{skyblue}$ Q_{ledq} $&$\color{red} Q_{qqq}$&$\color{red} Q_{duu}$ &$\color{red} Q_{qqu}$ & $\color{red} Q_{duq}$ \\
			\hline \hline
			&&15 &14 &13&13&13&13 &13&8&8&5&15&15&9&15&9&9 &5&5&5&13&13&13&13&13&13 &13&13&13 &13&13&13&13&13&12&10&7&9&9 &9&2&2&1 &1& 2& 2& 1& 1\\
			\hline \hline
			\text{    } $\mathcal{S}$&(1,1,0) & HL & \xmark& \xmark&\xmark& \xmark& \xmark&  \xmark& \xmark& \xmark& HL & T & HL & HL & T &\xmark & \xmark &HL &HL &HL &\xmark &\xmark & \xmark& \xmark& \xmark& \xmark &\xmark &\xmark & \xmark &\xmark &\xmark &\xmark &\xmark &\xmark &\xmark &\xmark & \xmark& \xmark&\xmark &\xmark & \xmark&\xmark &\xmark & \xmark &\xmark & \xmark &\xmark & \xmark \\
			\hline \hline
			\text{     }$\mathcal{S}_{2}$&(1,1,2) & HH &HH & HH& HH & HH& HH& HH& \xmark& \xmark& \xmark & HH & HH & \xmark & HH &\xmark & \xmark &\xmark &\xmark &\xmark &HH &\xmark &HH &HH &HH &HH &T &HH &HH &HH &HH &HH &HH &HH &\xmark &\xmark &\xmark &\xmark &\xmark &\xmark &\xmark &\xmark &\xmark & \xmark &\xmark & \xmark &\xmark & \xmark \\
			\hline \hline
			\text{     }$\Delta$&(1,3,0) & T &HH &\xmark &\xmark &\xmark &\xmark & \xmark&HH &HH & HL & T & HL & HH & T &\xmark & \xmark &T &T &T &\xmark &HH &\xmark &\xmark &\xmark &\xmark &\xmark &\xmark &\xmark &\xmark &\xmark &\xmark &\xmark &\xmark &\xmark &HH &\xmark &\xmark &\xmark &\xmark &\xmark &\xmark &\xmark &\xmark &\xmark & \xmark &\xmark & \xmark \\
			\hline \hline
			\text{     }$\mathcal{H}_{2}$&(1,2,${-\frac{1}{2}}$) & HH & HH&HH &HH &HH &HH &HH &HH &HH & HH & HH & HH & HH & T &\xmark & \xmark &T &T &T &HH &HH &HH &HH &HH &HH & HH&HH & HH &T &HH &HH & HH&T &T & HH& HH&\xmark &\xmark &\xmark &T & T&\xmark &T &\xmark & \xmark &\xmark & \xmark \\
			\hline
			\text{      }$\Delta_{1}$&(1,3,1) & T &T & HH& HH& HH &HH &HH &HH &HH & HH & T & HH & HH & T &\xmark & \xmark &T &T &T & HH &HH &HH &HH &HH & HH& HH&HH & HH &HH &HH &HH & HH&HH & HH&HH & HH&\xmark & \xmark&\xmark &\xmark &\xmark &\xmark & \xmark &\xmark & \xmark &\xmark & \xmark \\
			\hline
			\text{     }$\Sigma$&(1,4,${\frac{1}{2}}$) & HH &HH & HH&HH & HH &HH &HH &HH & HH& HH & HH & HH & HH & HH & \xmark& \xmark &HH &HH &HH & HH &HH &HH &HH &HH & HH& HH&HH & HH &HH &HH &HH & HH&HH & HH& HH&HH &\xmark & \xmark&\xmark &\xmark &\xmark &\xmark & \xmark &\xmark & \xmark &\xmark & \xmark \\
			\hline \hline
			\text{     }$\varphi_{1}$&(3,1,${-\frac{1}{3}}$) & HH & HH& HH& HH & HH& HH& HH& \xmark& \xmark& \xmark & HH & HH & \xmark & HH & HH & HH &\xmark &\xmark &\xmark &HH &HH &HH &HH &HH & HH&HH & T&HH &HH &HH &HH &HH &HH &HH &T &\xmark &HH &HH &HH &\xmark &\xmark &\xmark &\xmark &T &T &T & T \\
			\hline
			\text{     }$\varphi_{2}$&(3,1,$-{\frac{4}{3}}$) & HH & HH& HH& HH & HH& HH& HH& \xmark& \xmark& \xmark & HH & HH & \xmark & HH &HH & HH &\xmark &\xmark &\xmark & HH &HH &HH &HH &HH &HH &HH &HH &T &HH &HH &HH &HH &HH &HH &HH &\xmark &HH &HH &HH &\xmark &\xmark &\xmark &\xmark &\xmark & T &\xmark & \xmark \\
			\hline \hline
			\text{     }$\Theta_{1}$&(3,2,$\frac{1}{6}$) & HH & HH&HH & HH& HH &HH & HH&HH &HH & \xmark & HH & HH & HH & HH &HH & HH &\xmark &\xmark &\xmark & HH &HH &HH &HH &HH &HH & HH&HH & HH &HH &HH &T & HH&HH & HH&HH & HH &HH & HH&HH &\xmark &\xmark &\xmark & \xmark &\xmark & \xmark &\xmark & \xmark \\
			\hline
			\text{     }$\Theta_{2}$&(3,2,$\frac{7}{6}$) & HH &HH &HH & HH& HH &HH &HH &HH &HH & \xmark & HH & HH & HH & HH &HH & HH &\xmark &\xmark &\xmark & HH &HH &HH &HH &HH & HH& HH&HH & HH &HH &T &HH & T&HH &HH & HH& HH&HH & HH& HH&\xmark &\xmark &\xmark & \xmark &\xmark & \xmark &\xmark & \xmark \\
			\hline
			\text{     }$\Omega$&(3,3,-$\frac{1}{3}$) & HH &HH & HH& HH& HH &HH &HH &HH & HH& \xmark & HH & HH & HH & HH & HH& HH &\xmark &\xmark &\xmark & HH &HH &HH &HH &HH &T & HH&HH & HH &HH &HH &HH & HH&HH & HH& T& HH&HH & HH& HH&\xmark &\xmark &\xmark & \xmark &T & \xmark &\xmark & \xmark \\
			\hline \hline
			\text{     }$\chi_{_1}  $&(6,3,$\frac{1}{3}$) & HH &HH & HH&HH & HH &HH &HH &HH &HH & \xmark & HH & HH & HH & HH & HH& HH &\xmark &\xmark &\xmark & T&T & HH &HH &HH &HH & HH&HH & HH &HH &HH &HH & HH&HH &HH & HH&HH &HH & HH&HH &\xmark &\xmark &\xmark & \xmark &\xmark & \xmark &\xmark & \xmark \\
			\hline
			\text{     }$\chi_{_2}  $&(6,1,$\frac{4}{3}$) & HH &HH &HH &HH & HH &HH & HH& \xmark& \xmark& \xmark & HH & HH & \xmark & HH & HH& HH &\xmark &\xmark &\xmark & HH& HH&T &HH &HH & HH& HH&HH & HH &HH &HH &HH & HH&HH &HH & \xmark&\xmark &HH & HH& HH&\xmark &\xmark &\xmark & \xmark &\xmark & \xmark &\xmark & \xmark \\
			\hline
			\text{     }$\chi_{_3} $&(6,1,-$\frac{2}{3}$) & HH &HH &HH &HH & HH &HH & HH&\xmark & \xmark& \xmark & HH & HH & \xmark& HH & HH& HH &\xmark &\xmark &\xmark & HH& HH& HH&T &HH & HH& HH&HH & HH &HH &HH &HH & HH&HH &HH &\xmark &\xmark &HH &HH & HH&\xmark &\xmark &\xmark & \xmark &\xmark & \xmark &\xmark & \xmark \\
			\hline
			\text{     }$\chi_{_4}  $&(6,1,$\frac{1}{3}$) & HH & HH&HH &HH & HH &HH & HH&\xmark &\xmark & \xmark & HH & HH & \xmark & HH & HH& HH &\xmark &\xmark &\xmark &T &T & HH& HH& T& HH& HH&HH & HH &HH &HH &HH & HH&HH &HH &\xmark &\xmark &T & HH& HH&T &\xmark & T& \xmark &\xmark & \xmark &\xmark & \xmark \\
			\hline
		\end{tabular}
	\end{adjustbox}
\end{table}

\section{Interplay between Observables and Operators}\label{sec:obs-ops}

Once the heavy scalar fields are integrated out, each of the considered BSM scenarios can be expressed in terms of  SMEFT. Now all these emergent effective theories are described by the SM DOFs and symmetries, and thus are ready to be adjudged in the light of the same set of experimental observables. We have pooled the relevant effective operators based on their contributions to the following observables\footnote{Though some of the operators do not contribute to our chosen set of experimental observables they may affect other processes, including rare ones, which are not included here. Thus we will call all of them ``observables''  to keep them on the same footing for the rest of our analysis.} and characteristics as \cite{Ellis:2018gqa,Dawson:2019clf,Baglio:2020oqu,Dawson:2020oco,DasBakshi:2020ejz}:
\begin{eqnarray}\label{eq:ObsEffOps-ewpolo}
\color{blue} \text{EWPO-LO} &\color{blue} : & \color{blue}  \{Q_{HD},  Q_{HWB}, Q_{Hq}^{(1)}, Q_{Hq}^{(3)}, Q_{Hl}^{(1)}, Q_{Hl}^{(3)}, Q_{He}, Q_{Hu}, Q_{Hd}, Q_{ll}\}\color{blue};
\end{eqnarray}
\begin{eqnarray}\label{eq:ObsEffOps-ewponlo1} 
\color{beguni} \text{EWPO-NLO-I}  & \color{beguni} : &   \color{beguni} \{Q_{HB}, Q_{HW}, Q_{H\square}\} \color{beguni} ;
\end{eqnarray}
\begin{eqnarray} \label{eq:ObsEffOps-hss} 
\color{darkgreen}  \text{Higgs Signal Strength (HSS)} & \color{darkgreen} : & {\color{blue} \text{EWPO-LO} -\{Q_{ll}\}}+ {\color{beguni}\text{EWPO-NLO-I}} \nonumber\\ & & +{\color{darkgreen} \{ Q_{H}, Q_{uH}, Q_{dH}, Q_{eH}, Q_{G}, Q_{HG}\};}
\end{eqnarray}
\begin{eqnarray} \label{eq:ObsEffOps-ewponlo2} 
\color{gerua} \text{EWPO-NLO-II}  & \color{gerua} : & \color{gerua}  \{Q_{ed}, Q_{ee}, Q_{eu}, Q_{lu}, Q_{ld}, Q_{le},Q_{lq}^{(1)}, Q_{lq}^{(3)}, Q_{qe},   \nonumber \\
& & \color{gerua}  Q_{uB}, Q_{uW}, Q_{W}, Q_{qd}^{(1)}, Q_{qq}^{(1)}, Q_{qq}^{(3)}, Q_{qu}^{(1)}, Q_{ud}^{(1)}, Q_{uu}, Q_{dd}\};
\end{eqnarray}
\begin{eqnarray}\label{eq:ObsEffOps-adops} 
\color{skyblue} \text{Additional Operators (AdOps)} & \color{skyblue} : & \color{skyblue} \{Q_{ud}^{(8)}, Q_{qd}^{(8)}, Q_{qu}^{(8)}, Q_{quqd}^{(1)}, Q_{lequ}^{(1)}, Q_{quqd}^{(8)},  Q_{ledq}\}; 
\end{eqnarray}
\begin{eqnarray} \label{eq:ObsEffOps-blv}
\color{red} \text{B,L volating Operators (BLV)} & \color{red} : & \color{red} \{Q_{qqq}, Q_{duu}, Q_{qqu}, Q_{duq}\}.
\end{eqnarray}

We have tagged the operators with the same colour codes in Table~\ref{tab:BSMs-op-obs}. It is important to note that there are a few operators, [$Q_{ud}^{(8)}-Q_{ledq}$] in Table~\ref{tab:BSMs-op-obs}, that do not offer additional contributions to the observables of Eqs.~\ref{eq:ObsEffOps-ewpolo}-\ref{eq:ObsEffOps-blv}\,, while being generated from different BSM scenarios. Thus we have kept them under the label of ``Additional Operators'' (AdOps) and they are an integral part of our further analysis. On top of that, we find that $B, L$ violating operators emerge, see the last four columns in Table~\ref{tab:BSMs-op-obs}, which will play an important role in the model characterisation.

\noindent
This classification aims to show the explicit connection between observables and effective theories at the level of dimension-6 operators. It also highlights the large degree of overlapping contributions to the different observables, e.g. note the connection between HSS, EWPO-LO, and EWPO-NLO-I.  At this point, we need to consider the level of accuracy to which the effective operators are calculated. Is it sufficient to calculate the dimension-6 operators at the tree level or do we need to calculate operators at least up to 1-loop level? Also, can we break the degeneracy amongst the BSM theories using dimension-6 operators only, or do we have to look beyond? For example, if we restrict ourselves to integrating out heavy tree-level (T) propagators only, we will miss out on most of the effective operators, as a large number of SMEFT operators can be generated only at the loop level \cite{Grojean:2013kd,deBlas:2017xtg,Dawson:2020oco}. Thus, the operator computation up to tree-level may be misleading in drawing any conclusive remark about the BSM scenario, in particular in data-driven searches for new physics. It is also evident from Table~\ref{tab:BSMs-op-obs} that the higher-order corrections to the observables are sensitive to the loop induced effective operators. In this context, we further note that the additional operators that affect only the Higgs signal strengths appear only at loop level for colored scalars. Thus, unless we improve our level of precision in the computation of effective operators we will ignore the relevant constraints from precision measurements, which will lead to a reduced sensitivity in global parameter fits \cite{Ellis:2018gqa,Dawson:2019clf,Murphy:2017omb}.

\section{Observables and BSM classification}\label{sec:obs-bsm-class}

As discussed in the previous section, we want to adjudge the status of the BSM scenarios based on their response towards the different disjoint sets of effective dimension-6 operators. These operator sets are broadly defined by whether they (i) contribute to a specific set of observables, (ii) additional operators which do currently not affect our set of observables but potentially other observables, or whether they (iii) violate baryon or lepton number and lead to rare processes. To keep them on the same footing in the latter part of the discussion, we will denote all of them as ``observables'', as outlined in  Eqs.~\ref{eq:ObsEffOps-ewpolo}-\ref{eq:ObsEffOps-blv}.  

\vskip 0.5cm
The UV model classification proceeds according to the following steps:
\begin{itemize}
	
	\item First, we consider one ``observable'' at a time and thereby the corresponding set of contributing operators, as outlined in Eqs.~\ref{eq:ObsEffOps-ewpolo}-\ref{eq:ObsEffOps-blv}.

\item Then, we tag each of the BSM theories depending on whether they induce the respective operator following the results of Table~\ref{tab:BSMs-op-obs}. 

\item We further inspect each model ($M_i$) by computing in how far their operator set overlaps with that of other models, i.e. $(M_i - M_j)$ 
$\forall\; i \neq j$. Then we pool multiple BSMs together iff $(M_i - M_j)=\emptyset$ $\forall\; i \neq j$ and declare them to be degenerate for the sake of our model selection criteria.
This then also clarifies which models have a non-overlapping operator(s), which may lead to smoking-gun features for those BSM scenarios. 

\item For each of the ``observables'' we continue to perform these analysis steps to identify their individual impact on the latter part of the analysis.

\item At the end we will discuss the cumulative effect of all ``observables'' on the BSM model classification. We will also highlight the degeneracies in the model space if any are left, and we will use this result to motivate beyond the set of dimension-6 operators.

\end{itemize}

\subsection{EWPO-LO and BSM classification}\label{sec:EWPO-LO}

\begin{figure}[htb]
	\centering
	\includegraphics[height=8cm,width=11cm]{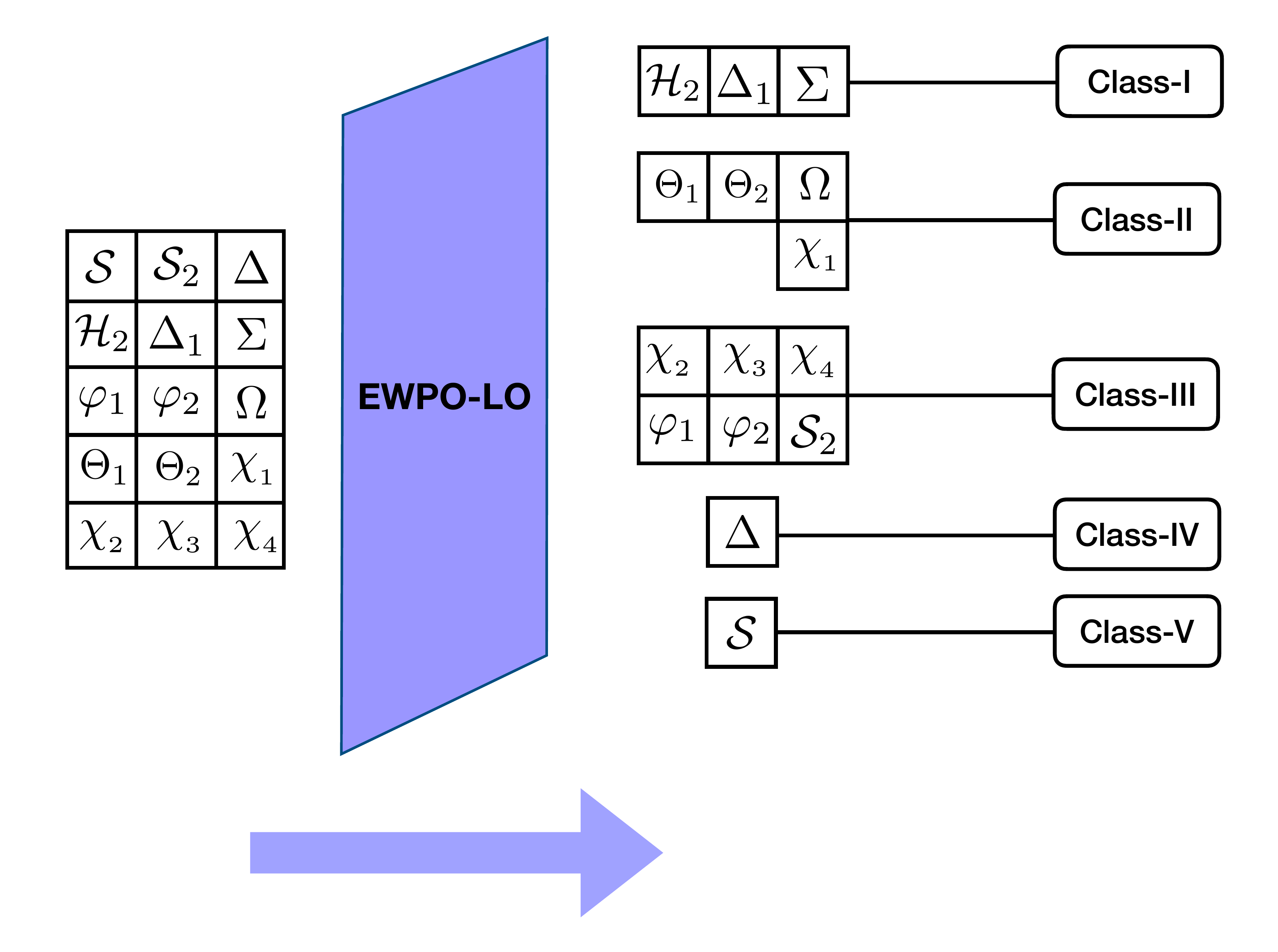}
	\caption{The BSM classification in the light of effective operators affecting EWPO-LO.}
	\label{fig:degeneracy-breaking-EWPO-LO}
\end{figure}

The dimension-6 effective operators $\color{blue}  \{Q_{HD},  Q_{HWB}, Q_{Hq}^{(1)}, Q_{Hq}^{(3)}, Q_{Hl}^{(1)}, Q_{Hl}^{(3)}, Q_{He}, Q_{Hu}, Q_{Hd}, Q_{ll}\}\color{blue}$ provide additional contributions to the electro-weak precision observables at leading order (EWPO-LO), see Eq.~\ref{eq:ObsEffOps-ewpolo}.
However, not all the operators are generated simultaneously once the heavy scalar field that belongs to one of the 15 BSM scenarios is integrated out, see Table~\ref{tab:BSMs-op-obs}. Thus from the perspective of these operators not all the BSM models are not on the same footing and can thus be classified as
\begin{eqnarray}\label{eq:op-class-EWPO-LO}
\text{Class-I:} & \Phi_i^{\text{I}} \in \{ \mathcal{H}_2, \Delta_1, \Sigma\} &\Rightarrow  \text{10 operators}; \nonumber\\
\text{Class-II:} &  \Phi_i^{\text{II}} \in \{ \Theta_1, \Theta_2, \Omega,\chi_{_1}\}& \Rightarrow  \text{9 operators}; \nonumber\\
\text{Class-III:} &  \Phi_i^{\text{III}} \in \{\mathcal{S}_2,  \phi_1, \phi_2, \chi_{_2},\chi_{_3},\chi_{_4} \}& \Rightarrow  \text{7 operators}; \\
\text{Class-IV:} &  \{ \Delta\} &\Rightarrow \text{5 operators};  \nonumber \\
\text{Class-V:} & \{ \mathcal{S}\} & \Rightarrow \text{2 operators}.  \nonumber
\end{eqnarray}
The status of these effective theories derived from different BSMs are captured in Fig.~\ref{fig:degeneracy-breaking-EWPO-LO}.
We find that the 15 BSM theories can be pooled into five classes I-V and amongst them the models belonging to classes I-III are degenerate, i.e. 
\begin{equation}\label{eq:LO-degeneracy}
\Phi_i^\text{I}- \Phi_j^\text{I} =\Phi_i^\text{II} - \Phi_j^\text{II}=\Phi_i^\text{III} - \Phi_j^\text{III}=\emptyset, \; \forall\; i,j. 
\end{equation}

\begin{figure}
	\includegraphics[height=5cm,width=15cm]{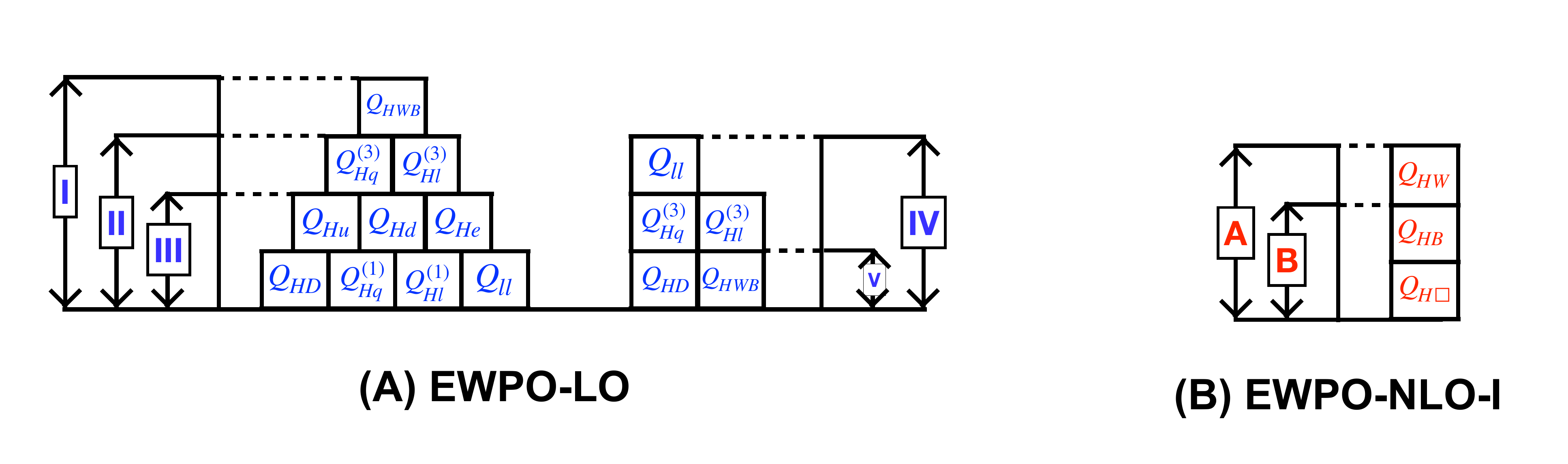}
	\caption{Distribution of effective operators affecting (A) EWPO-LO, (B) EWPO-NLO-I among different BSM classes.}
	\label{fig:BSMs-op-obs}
\end{figure}

Here, the maximum and minimum number of operators are contained in models within Class-I and Class-V respectively
\begin{equation}\label{eq:max-min-diff-LO-I}
\text{I} - \text{V} =  {\color{blue}  \{Q_{Hq}^{(1)}, Q_{Hq}^{(3)}, Q_{Hl}^{(1)}, Q_{Hl}^{(3)}, Q_{He}, Q_{Hu}, Q_{Hd}, Q_{ll}\}},
\end{equation}
and the operators of all other classes always form a subset of
\begin{equation}\label{eq:class-diff-LO-I}
\text{I} - \text{IV}={\color{blue}\{Q_{Hq}^{(1)}, Q_{Hl}^{(1)}, Q_{He}, Q_{Hu}, Q_{Hd}\}} ;~
\text{I} - \text{III}= { \color{blue}\{Q_{Hq}^{(3)},  Q_{Hl}^{(3)}\}}; ~\text{I} - \text{II} = { \color{blue} \{Q_{HWB}\}}.
\end{equation}
The pivotal role of Class-I and the various subsets of operators have been depicted in the operator-pyramids of Fig.~\ref{fig:BSMs-op-obs}(A).  It also shows that all the observables affected by models of classes II-V are however also affected by models belonging to  Class-I. Thus, models belonging to classes II-V are lacking any smoking gun features to separate them from Class-I models.

\subsection{EWPO-NLO-I and BSM classification}\label{sec:EWPO-NLO-I}

\begin{figure}[htb!]
	\centering
	\includegraphics[height=8cm,width=11cm]{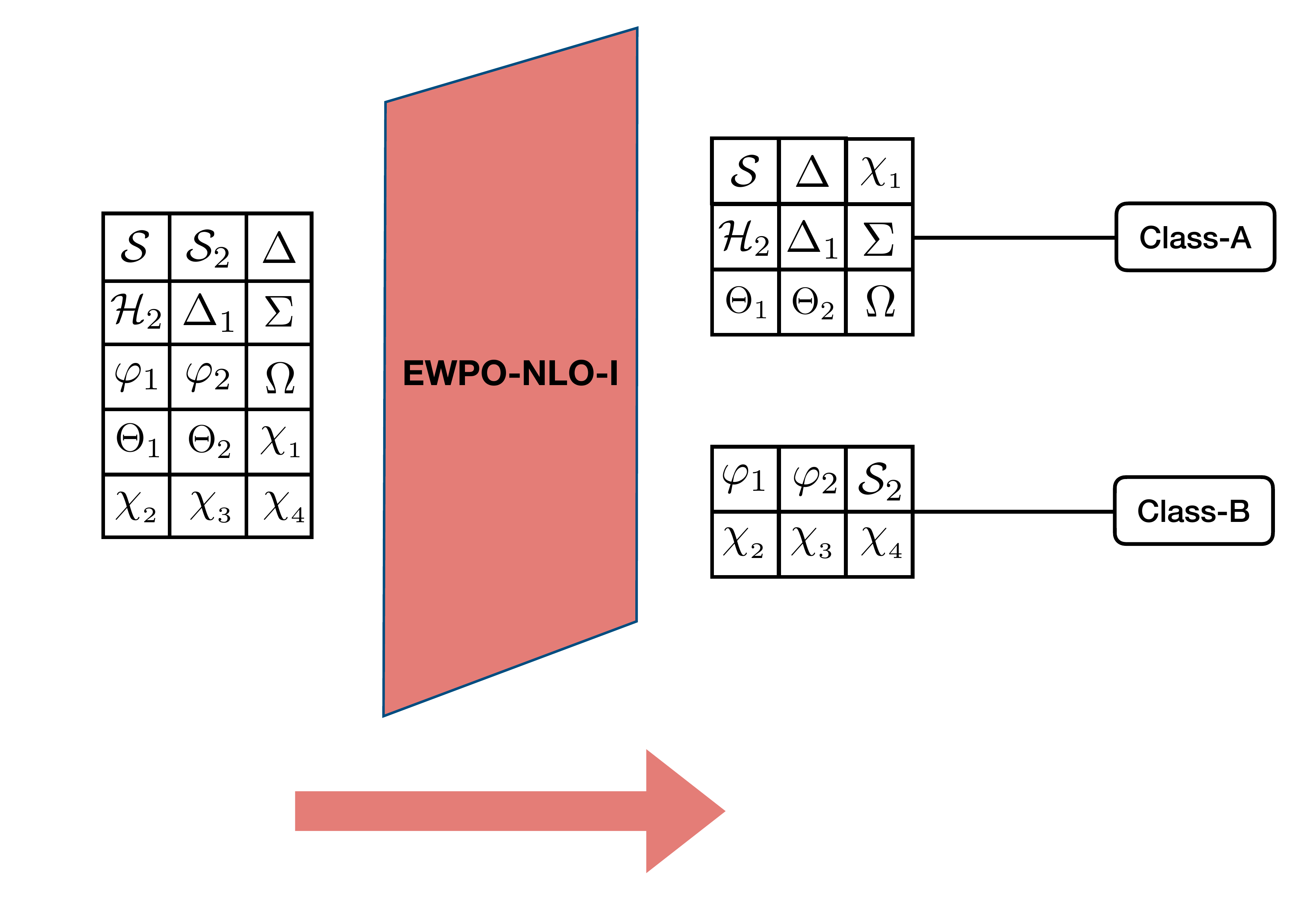}
	\caption{The BSM classification in the light of effective operators affecting EWPO-NLO-I.}
	\label{fig:degeneracy-breaking-EWPO-NLO-I}
\end{figure}

The operators of Eq.~\ref{eq:ObsEffOps-ewponlo1}, $ \color{beguni} \{Q_{HB}, Q_{HW}, Q_{H\square}\} \color{beguni} $ contribute to electro-weak precision observables at 1-loop level, i.e. EWPO-NLO-I.
All 15 BSM scenarios fall into two categories in relation to these operators
\begin{eqnarray}\label{eq:op-class-EWPO-NLO-I}
\text{Class-A:} & \Phi_i^{\text{A}} \in \{\mathcal{S},\Delta, \mathcal{H}_2, \Delta_1, \Sigma, \Theta_1, \Theta_2, \Omega,\chi_{_1}\} &\Rightarrow  \text{3 operators};\nonumber\\
\text{Class-B:} &  \Phi_i^{\text{B}} \in \{\mathcal{S}_2,\phi_1,\phi_2,\chi_{_2},\chi_{_3},\chi_{_4}\} &\Rightarrow  \text{2 operators}.
\end{eqnarray}
In Fig.~\ref{fig:degeneracy-breaking-EWPO-NLO-I}, we have captured the degeneracy in the model space with respect to these three effective operators $ \color{beguni} \{Q_{HB}, Q_{HW}, Q_{H\square}\} \color{beguni} $ only. 

Similar to the discussion in Sec.~\ref{sec:EWPO-LO}, the status of these classes and the distribution of operators are shown in Fig.~\ref{fig:BSMs-op-obs}(B). Here, Class-A BSMs contain all  three operators and comprise all the ``effective'' features of Class-B scenarios 
\begin{equation}\label{eq:max-min-diff-NLO-I}
\text{A} - \text{B} =  {\color{beguni}  \{Q_{HW}\} }.
\end{equation}

\subsection{Higgs Signal Strengths and BSM classification}\label{sec:Higgs-SS}

\begin{figure}[hb]
	\centering
		\includegraphics[height=5cm,width=10cm]{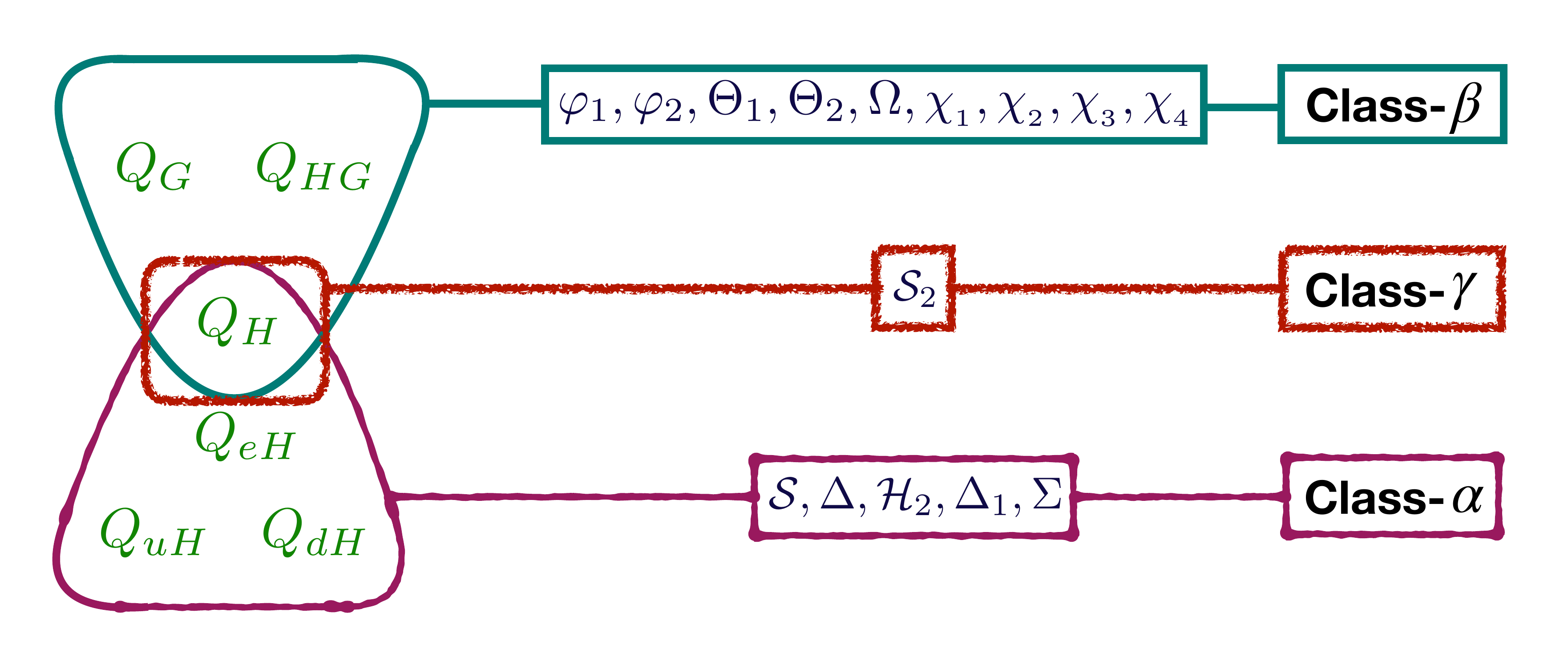}
		\caption{Status of BSMs {\it w.r.to.} the following operator set: ${\color{darkgreen} \{Q_{H}, Q_{uH}, Q_{dH}, Q_{eH}, Q_{G}, Q_{HG}\}}$.}
	\label{fig:ops-class-Higgs}
\end{figure}

The effective operators that affect the Higgs signal strengths also contribute to the EWPO-LO and EWPO-NLO-I sets. There are six additional operators $ {\color{darkgreen} \{Q_{H}, Q_{uH}, Q_{dH}, Q_{eH}, Q_{G}, Q_{HG}\}}$ that are exclusive to the HSS. After adjudging the status of the BSMs using EWPO-LO and EWPO-NLO-I, here, we use  these six operators to classify the BSMs as
\begin{eqnarray}\label{eq:op-class-Higgs4ops}
	\text{Class-$ \alpha $:} & \Phi_i^{\alpha} \in \{ \mathcal{S}, \Delta, \mathcal{H}_2, \Delta_{1}, \Sigma\} &\Rightarrow   {\color{darkgreen} \{Q_{H}, Q_{uH}, Q_{dH}, Q_{eH} \}};\nonumber\\
	\text{Class-$ \beta $:} &  \Phi_i^{\beta} \in \{\varphi_1, \varphi_2, \Theta_1, \Theta_2, \Omega,\chi_{_1}, \chi_{_2},\chi_{_3},\chi_{_4} \} &\Rightarrow  {\color{darkgreen} \{Q_{H}, Q_{G}, Q_{HG} \}};\\
	\text{Class-$ \gamma $:} &  \{ \mathcal{S}_2\} &\Rightarrow  {\color{darkgreen} \{Q_{H}\}}.\nonumber
\end{eqnarray}
We further observe that two classes  $\alpha$ and $\beta$ contain multiple degenerate BSM scenarios, see  Eq.~\ref{eq:op-class-Higgs4ops},
\begin{equation}\label{}
\Phi_i^\alpha - \Phi_j^\alpha =\Phi_i^\beta - \Phi_j^\beta= \emptyset, \; \forall\; i,j. 
\end{equation}
Unlike to the cases discussed previously in Secs.~\ref{sec:EWPO-LO} and \ref{sec:EWPO-NLO-I}, here, none of the BSM scenarios induces all of the six effective operators simultaneously, see Eq.~\ref{eq:op-class-Higgs4ops}. Thus the operator sets lack the earlier pyramidal structure and rather form a semi-disjoint convoluted pattern, see Fig.~\ref{fig:ops-class-Higgs}.
\vskip 0.2cm

At this point, it is now possible to classify the UV models based on the cumulative impact of EWPO-LO, EWPO-NLO-I, and the HSS as
\begin{align*}\label{eq:op-class-Higgs-SS}
	\text{Class-H1:} \Phi_i^{\text{H1}} \in \{ \mathcal{H}_2, \Delta_1, \Sigma\} \Rightarrow & \{  {\color{darkgreen}Q_{HD},  Q_{HWB}, Q_{Hq}^{(1)}, Q_{Hq}^{(3)}, Q_{Hl}^{(1)}, Q_{Hl}^{(3)}, Q_{He}, Q_{Hu}, Q_{Hd}, \underline{Q_{ll}},}\\
	&{\color{darkgreen}Q_{HB}, Q_{HW}, Q_{H\square}, Q_{H}, Q_{uH}, Q_{dH}, Q_{eH}}\}\\
	\equiv & \text{{\color{blue}Class-I} + {\color{beguni}Class-A}}+\{{\color{darkgreen}Q_{H}, Q_{uH}, Q_{dH}, Q_{eH}}\};\\
	\text{Class-H2:} \Phi_i^{\text{H2}} \in \{ \Theta_1, \Theta_2, \Omega,\chi_{_1}\} \Rightarrow & \{ {\color{darkgreen}Q_{HD},   Q_{Hq}^{(1)}, Q_{Hq}^{(3)}, Q_{Hl}^{(1)}, Q_{Hl}^{(3)}, Q_{He}, Q_{Hu}, Q_{Hd}, \underline{Q_{ll}},}\\
	&{\color{darkgreen}Q_{HB}, Q_{HW}, Q_{H\square}, Q_{H},  Q_{G}, Q_{HG}\}}\\
	\equiv & \text{{\color{blue}Class-II} + {\color{beguni}Class-A}}+\{{\color{darkgreen}Q_{H}, Q_{G}, Q_{HG}}\};\\
	\text{Class-H3:} \Phi_i^{\text{H3}} \in \{\phi_1, \phi_2, \chi_{_2},\chi_{_3},\chi_{_4} \} \Rightarrow & \{  {\color{darkgreen}Q_{HD},   Q_{Hq}^{(1)},  Q_{Hl}^{(1)}, Q_{He}, Q_{Hu}, Q_{Hd}, \underline{Q_{ll}},}\nonumber\\
	&{\color{darkgreen}Q_{HB}, Q_{H\square}, Q_{H},  Q_{G}, Q_{HG}}\};\nonumber\\
	\equiv & \text{{\color{blue}Class-III} + {\color{beguni}Class-B}}+\{{\color{darkgreen}Q_{H}, Q_{G}, Q_{HG}}\};
\end{align*}
\vspace{-1cm}
\begin{align}
	\text{Class-H4:} \{ \Delta\} \Rightarrow & \{{\color{darkgreen}Q_{HD},  Q_{HWB}, Q_{Hq}^{(3)}, Q_{Hl}^{(3)}, \underline{Q_{ll}}, Q_{HB}, Q_{HW}, Q_{H\square}, Q_{H}, Q_{uH}, Q_{dH}, Q_{eH}} \} \nonumber\\
	\equiv & \text{{\color{blue}Class-IV} + {\color{beguni}Class-B}}+\{{\color{darkgreen}Q_{H}, Q_{uH}, Q_{dH}, Q_{eH}}\};\nonumber\\
	\text{Class-H5:} \{ \mathcal{S}_2\} \Rightarrow & \{ {\color{darkgreen}Q_{HD},   Q_{Hq}^{(1)}, Q_{Hl}^{(1)}, Q_{He}, Q_{Hu}, Q_{Hd}, \underline{Q_{ll}}, Q_{HB}, Q_{H\square}, Q_{H}} \} \nonumber\\
	\equiv & \text{{\color{blue}Class-III} + {\color{beguni}Class-B}}+\{{\color{darkgreen}Q_{H}}\};\nonumber\\
	\text{Class-H6:} \{ \mathcal{S}\} \Rightarrow & \{{\color{darkgreen}Q_{HD},  Q_{HWB}, Q_{HB}, Q_{HW}, Q_{H\square}, Q_{H}, Q_{uH}, Q_{dH}, Q_{eH}}\}\nonumber\\
	\equiv & \text{{\color{blue}Class-V} + {\color{beguni}Class-A}}+\{{\color{darkgreen}Q_{H}, Q_{uH}, Q_{dH}, Q_{eH}}\}.
\end{align}

\begin{figure}[htb!]
	\centering
	\includegraphics[height=8cm,width=11cm]{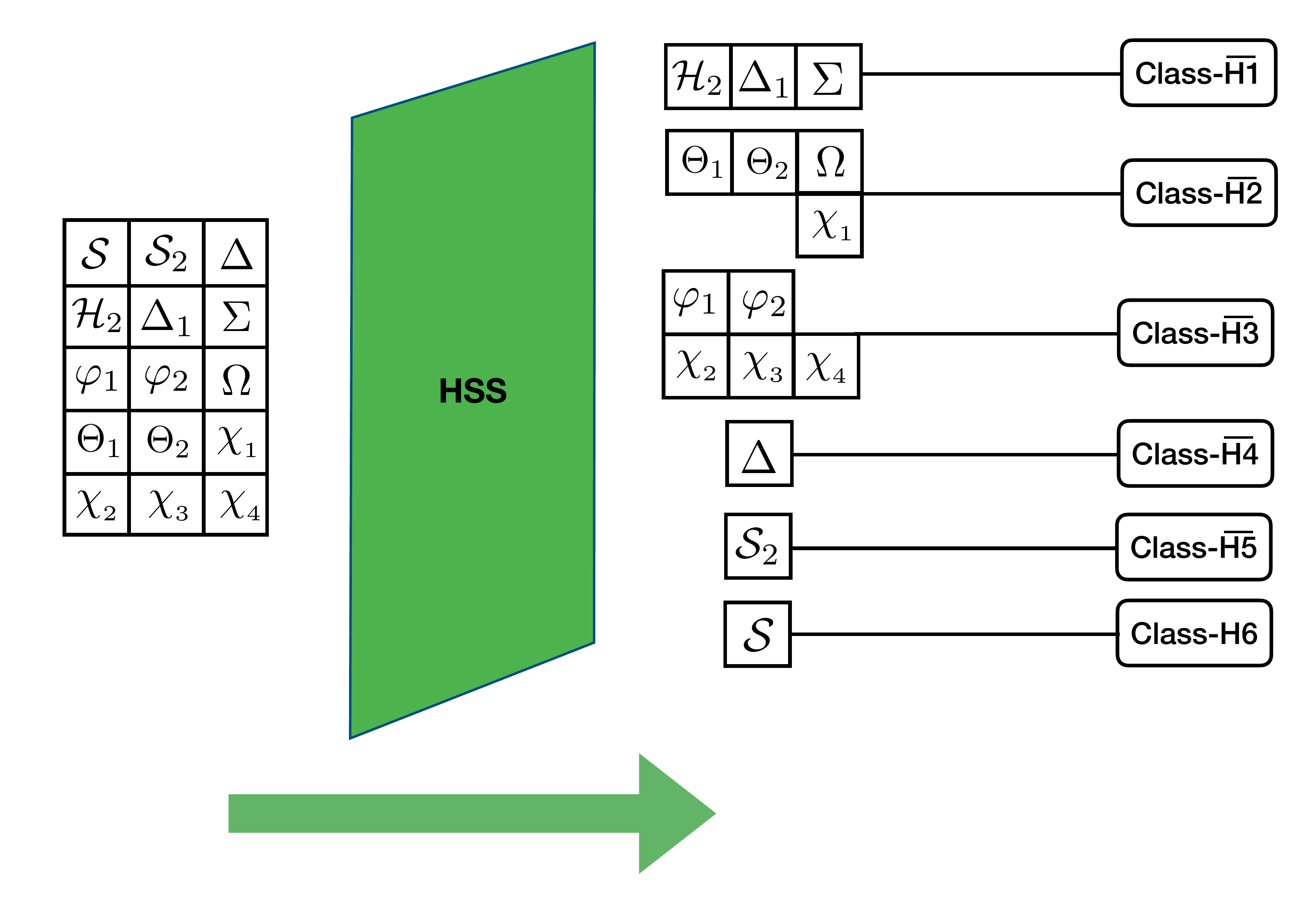}
	\caption{The BSM classification in the light of effective operators affecting Higgs Signal Strength (HSS). The bars on Hi's signifies the absence of  $ Q_{ll} $ from HSS effective operator set.}
	\label{fig:degeneracy-breaking-Higgs-data}
\end{figure}

Here, we find that there are three degenerate classes H1, H2, and H3 which contain multiple BSM models, see Eq.~\ref{eq:op-class-Higgs-SS}, and satisfy
\begin{equation}\label{}
	\Phi_i^\text{H1} - \Phi_j^\text{H1} =\Phi_i^\text{H2} - \Phi_j^\text{H2} =\Phi_i^\text{H3} - \Phi_j^\text{H3}=\emptyset, \; \forall\; i,j. 
\end{equation}
We further identify that the maximum and the minimum number of operators are contained within Classes-H1 and -H6 respectively, and the distribution of the operators among them is convoluted, see the earlier discussions. 
The impact of Higgs data on the BSM classification can be easily understood from Eq.~\ref{eq:op-class-Higgs-SS} which has been captured in  Fig.~\ref{fig:degeneracy-breaking-Higgs-data}, where  $\overline{\text{Hi}}$ is defined as $(\text{Hi}-Q_{ll})$ $\forall\;\text{i}$, see Eq.~\ref{eq:op-class-Higgs-SS}.

So far, we have discussed the individual and cumulative impact of different ``observables'' on sets of operators induced by the various scalar BSM scenarios. This indicates the importance of a suitable choice of ``observables'' for a comprehensive interpretation in terms of new physics models and in turn allows to prioritise measurements in their relevant importance to uniquely identify UV models. From Fig.~\ref{fig:degeneracy-breaking-Higgs-data} it is evident that even after the inclusion of the above set of measurements, we are left with a few indistinguishable classes, e.g., $\overline{\text{H1}}, \overline{\text{H2}},\overline{\text{H3}}$, which contain multiple models that generate the same effective dimension-6 operators.

\subsection{EWPO-NLO-II and BSM classification}\label{sec:EWPO-NLO-II}

\begin{figure}
	\centering
	\includegraphics[height=5cm,width=8cm]{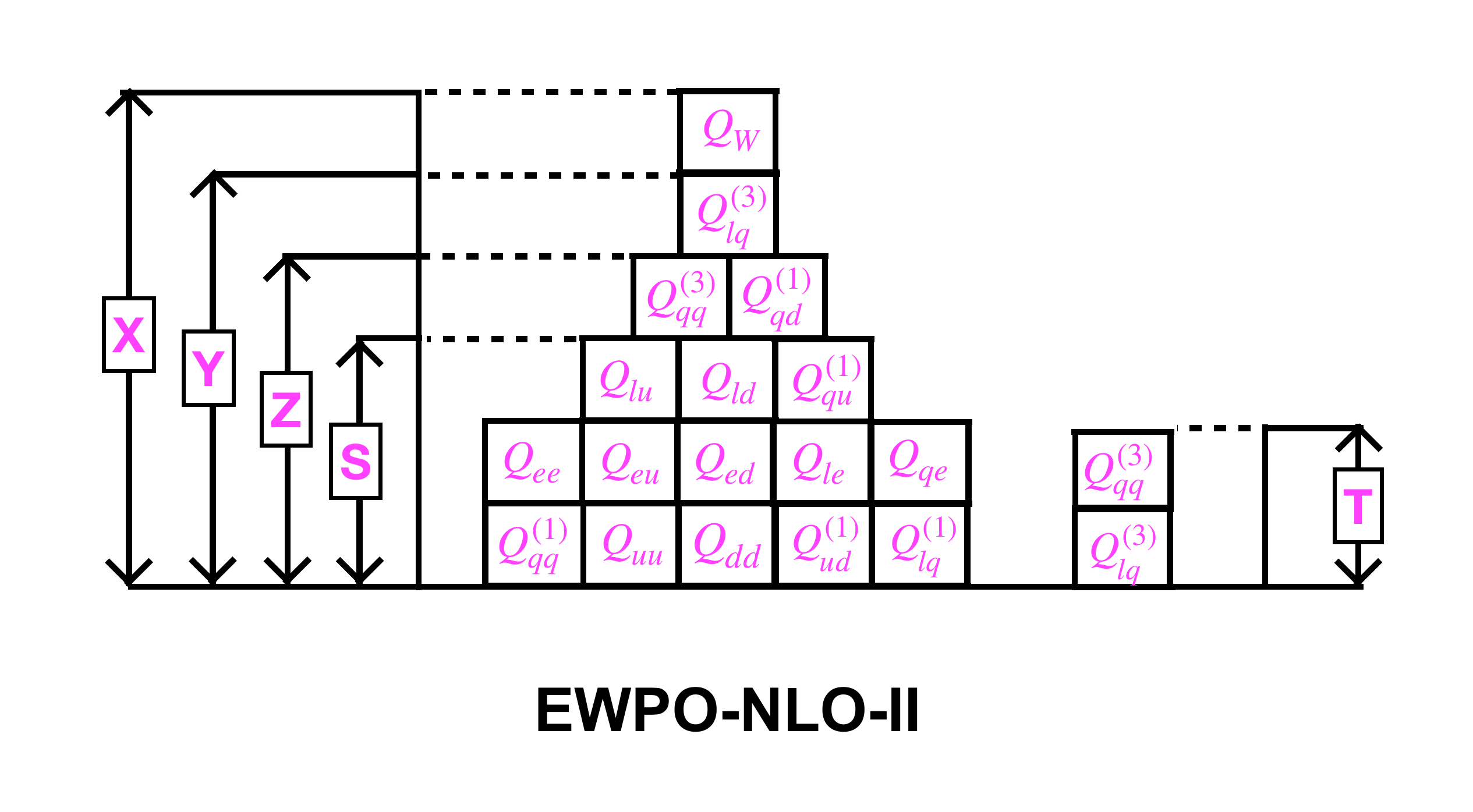}
	\caption{Distribution of effective operators affecting EWPO-NLO-II among different BSM classes.}
	\label{fig:BSMs-op-obs-EWPO-NLO-II}
\end{figure}

19 dimension-6 operators $  \color{gerua}  \{Q_{ed}, Q_{ee}, Q_{eu}, Q_{lu}, Q_{ld}, Q_{le},Q_{lq}^{(1)}, Q_{lq}^{(3)}, Q_{qe},Q_{uB}, Q_{uW}, Q_{W}, Q_{qd}^{(1)}, Q_{qq}^{(1)}, Q_{qq}^{(3)},\\ Q_{qu}^{(1)}, Q_{ud}^{(1)}, Q_{uu}, Q_{dd}\}$ contribute to the observables EWPO-NLO-II, see Eq.~\ref{eq:ObsEffOps-ewponlo2}. However, two those operators $\color{gerua} Q_{uB}, Q_{uW}$ do not emerge from the BSM models when integrating out the heavy degree of freedom up to 1-loop level, see Table~\ref{tab:BSMs-op-obs}. Thus we are left with seventeen effective operators that affect EWPO-NLO-II. Consequently, the BSM theories can be categorised as 
\begin{eqnarray}\label{eq:op-class-EWPO-NLO-II-SS}
\text{Class-X:} & \Phi_i^{\text{X}} \in \{ \mathcal{H}_2, \Delta_{1}, \Sigma, \Theta_1, \Theta_2, \Omega, \chi_{_1}\} &\Rightarrow  \text{17 operators};\nonumber\\
\text{Class-Y:} &  \Phi_i^{\text{Y}} \in \{\varphi_1, \varphi_2\} &\Rightarrow  \text{16 operators};\nonumber\\
\text{Class-Z:} &  \Phi_i^{\text{Z}} \in \{\chi_{_2},\chi_{_3},\chi_{_4}\} &\Rightarrow \text{15 operators};\\
\text{Class-S:} &  \{\mathcal{S}_2\} &\Rightarrow \text{13 operators};\nonumber\\
\text{Class-T:} &   \{\Delta\} &\Rightarrow  \text{2 operators};\nonumber\\
\text{Class-U:} &   \{\mathcal{S}\} &\Rightarrow  \text{0 operator}\nonumber.
\end{eqnarray}

Here, we find three degenerate classes X, Y, and Z containing multiple BSM scenarios
\begin{equation}\label{}
\Phi_i^\text{X} - \Phi_j^\text{X} =\Phi_i^\text{Y} - \Phi_j^\text{Y}=\Phi_i^\text{Z} - \Phi_j^\text{Z}=\emptyset, \; \forall\; i,j. 
\end{equation}
The maximum and the minimum number of operators are contained within Class-X and Class-U respectively, and the operator sets form a pyramidal structure similar to the EWPO-LO and EWPO-NLO-I cases, see Fig.~\ref{fig:BSMs-op-obs-EWPO-NLO-II}.
Thus all other operator sets belonging to Y, Z, T, and U are contained within Class-X
\begin{equation*}\label{eq:class-diff-NLO-II}
\text{X} - \text{Y}={\color{gerua}\{     Q_{W}\}} ; \text{X} - \text{Z} = {\color{gerua}\{Q_{lq}^{(3)}, Q_{W}\}}; \text{X} - \text{S} = {\color{gerua}  \{ Q_{qd}^{(1)},  Q_{qq}^{(3)}, Q_{lq}^{(3)}, Q_{W}\}};\text{X} - \text{U} = \text{X};
\end{equation*}
\begin{equation}
\text{X} - \text{T} = {\color{gerua}  \{Q_{ed}, Q_{ee}, Q_{eu}, Q_{lu}, Q_{ld}, Q_{le},Q_{lq}^{(1)},  Q_{qe}, Q_{W}, Q_{qd}^{(1)}, Q_{qq}^{(1)},  Q_{qu}^{(1)}, Q_{ud}^{(1)}, Q_{uu}, Q_{dd}\}}.
\end{equation}
This indicates that Class-X is superior to others and contains all the features from other classes in this classification.
\begin{figure}
	\centering
	\includegraphics[height=8cm,width=11cm]{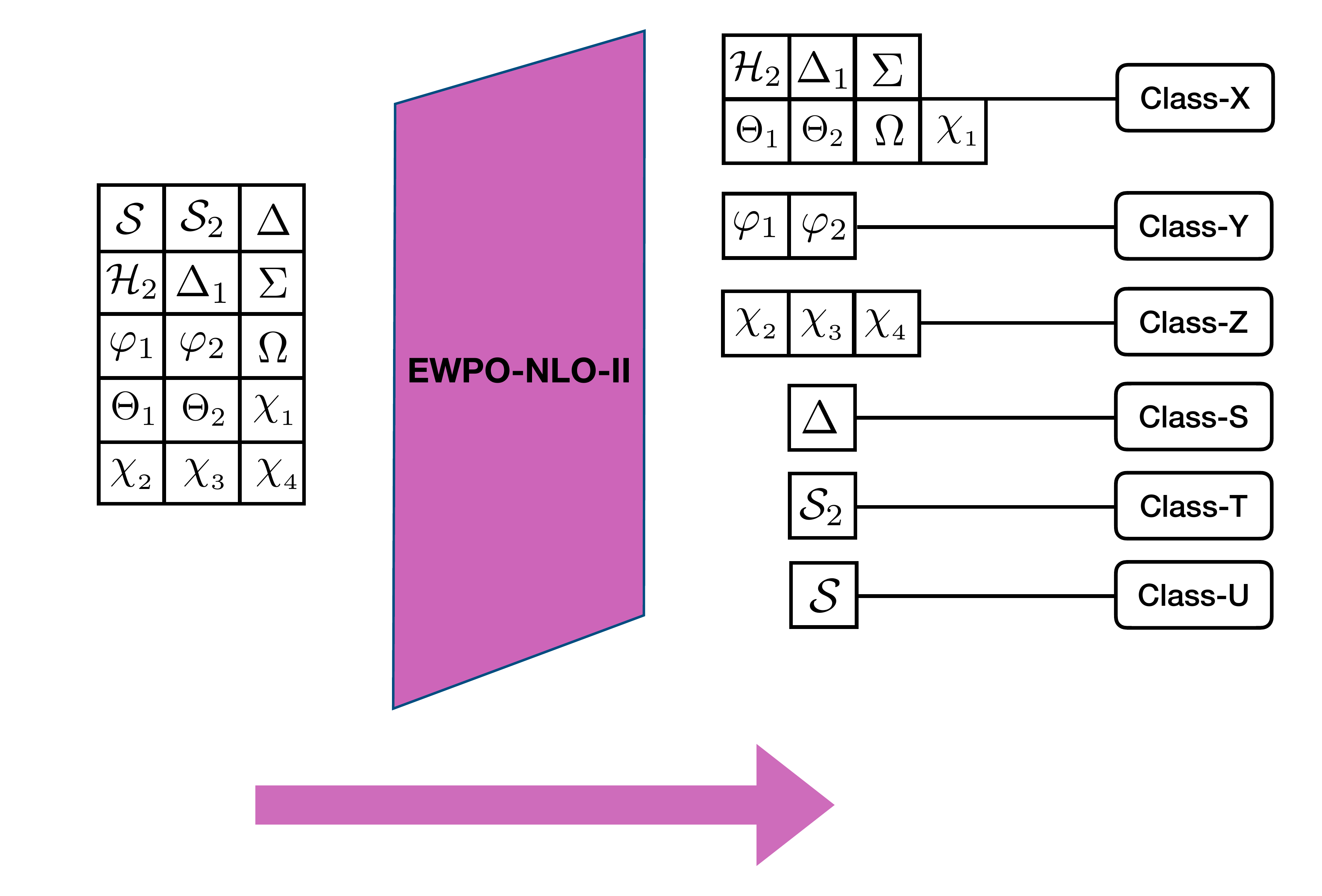}
	\caption{The BSM classification in the light of effective operators affecting EWPO-NLO-II.}
	\label{fig:degeneracy-breaking-EWPO-NLO-II}
\end{figure}
In Fig.~\ref{fig:degeneracy-breaking-EWPO-NLO-II}, we show the BSM classification and highlight the degeneracy in model space with respect to these seventeen effective operators:

\subsection{Additional Operators and BSM classification}

\begin{figure}
	\centering
	\includegraphics[height=8cm,width=11cm]{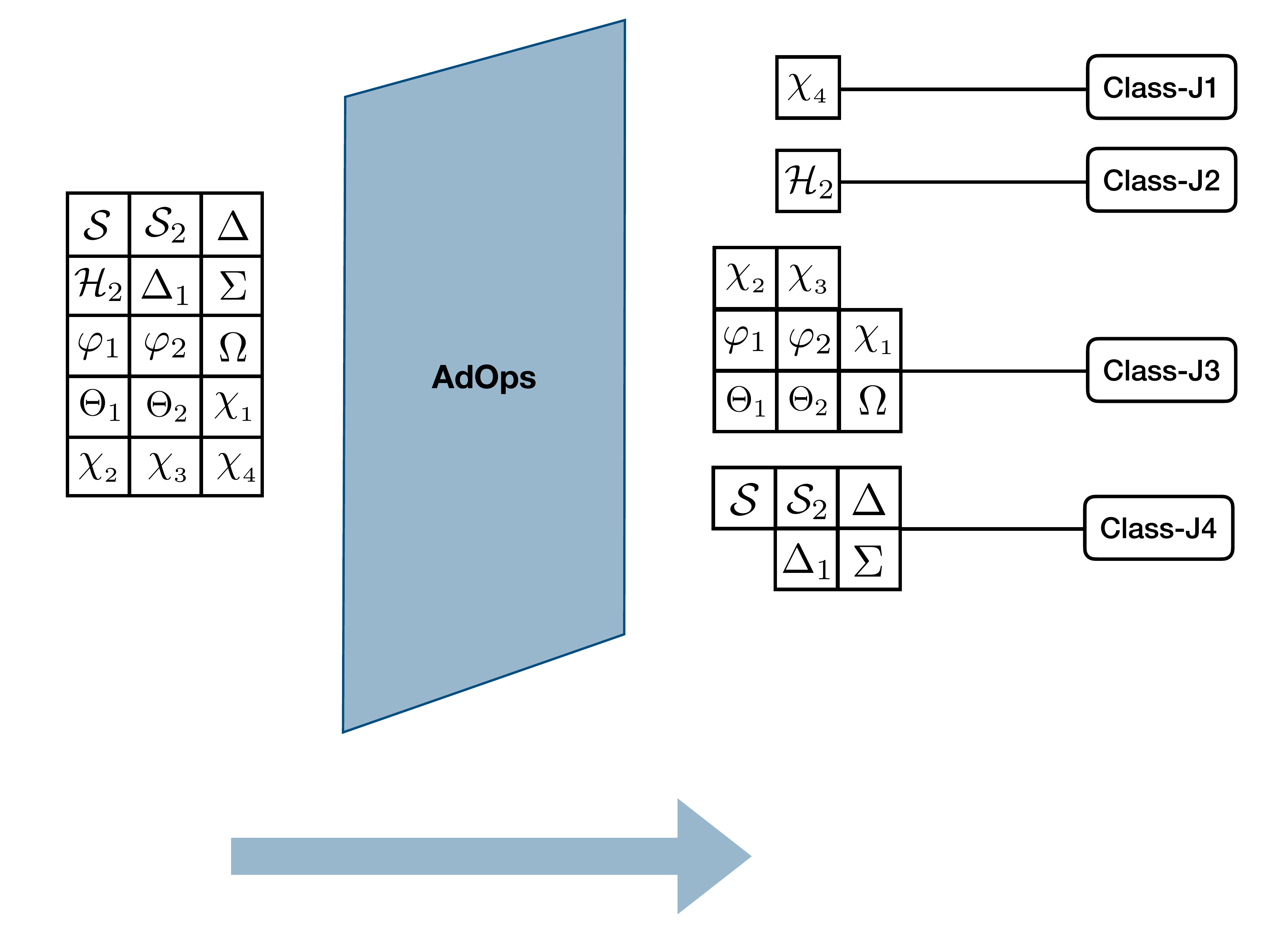}
	\caption{The BSM classification in the light of effective operators affecting Additional Operators.}
	\label{fig:degeneracy-breaking-Add-Ops}
\end{figure}

Some of the operators do not contribute to the list of our experimental observables discussed so far, but do emerge in the process of integrating out the heavy fields belonging to the example BSM scenarios. The relation of the operators $\{{\color{skyblue} Q_{ud}^{(8)}, Q_{qd}^{(8)}, Q_{qu}^{(8)}, Q_{quqd}^{(1)}, Q_{lequ}^{(1)}, Q_{quqd}^{(8)}, Q_{ledq}}\}$ to the respective BSM models can be captured as
\begin{eqnarray}\label{eq:op-class-Add-Ops}
	\text{Class-J1:} & \{\chi_{_4}\} &\Rightarrow \text{5 operators};\nonumber\\
	\text{Class-J2:} &  \{\mathcal{H}_2\} &\Rightarrow \text{3 operators};\nonumber\\
	\text{Class-J3:} &  \Phi_i^{\text{J3}} \in \{\varphi_1, \varphi_2, \Theta_1, \Theta_2, \Omega, \chi_{_1}, \chi_{_2}, \chi_{_3}\}&\Rightarrow  \text{3 operators};\\
	\text{Class-J4:} &  \Phi_i^{\text{J4}} \in  \{ \mathcal{S}, \mathcal{S}_2, \Delta, \Delta_{1}, \Sigma\} &\Rightarrow  \text{0 operator}\;.\nonumber
\end{eqnarray}
We find two degenerate classes, J3 and J4,  containing multiple BSM models
\begin{equation}\label{}
	\Phi_i^\text{J3} - \Phi_j^\text{J3} =\Phi_i^\text{J4} - \Phi_j^\text{J4}=\emptyset, \; \forall\; i,j. 
\end{equation}
Again, the relation between the effective operators and the BSM models is not of pyramidal structure. Although the majority of operators are contained in Class-J1, this class does not contain the features of many UV models. The BSM classification is displayed in Fig.~\ref{fig:degeneracy-breaking-Add-Ops}.

\subsection{$B,L$ violation: importance of rare process}
All the operators that have been taken into consideration so far preserve $B$ and $L$ numbers. However, in three of the BSM models $(\phi_1,\phi_2,\chi_{_4})$ four $B,L$ violating dimension-6 operators $ \{{\color{red} Q_{qqq}, Q_{duu}, Q_{qqu}, Q_{duq}}\} $ are generated once the heavy fields are integrated out, see Table~\ref{tab:BSMs-op-obs}.
These $B,L$ violating operators may give rise to rare processes which have essentially no SM background, and can thus be considered smoking-gun features of the new physics scenarios. As few BSM scenarios induce such operators, any observed measurement induced by them will signify the presence of interactions that involve any of the particles $\phi_1,\phi_2$ or $\chi_{_4}$.

\section{BSM Classification: present and future}\label{sec:BSM-class-present-future}

We have shown how the 15 new physics models introduced in Sec.~\ref{sec:obs-bsm-class} can be distinguished from each other experimentally, based on the different operators they introduce and how they affect precision observables. Limiting our analysis to dimension-6 operators, we were able to prepare a map that shows how the different observables and theory-imposed operator hierarchies break the degeneracies between the various BSM scenarios. 
In the process, we have identified classes of models that reflect that the models within such a class induce the same operator and affect the observable at hand in the same way. By scrutinising the inner structures of different classes to understand the distribution of operators among them and that in turn helps us to identify the non-overlapping operators. This labeling will help us to design the smoking gun features of the individual class of models. 

\begin{figure}
	\includegraphics[height=10cm,width=18cm]{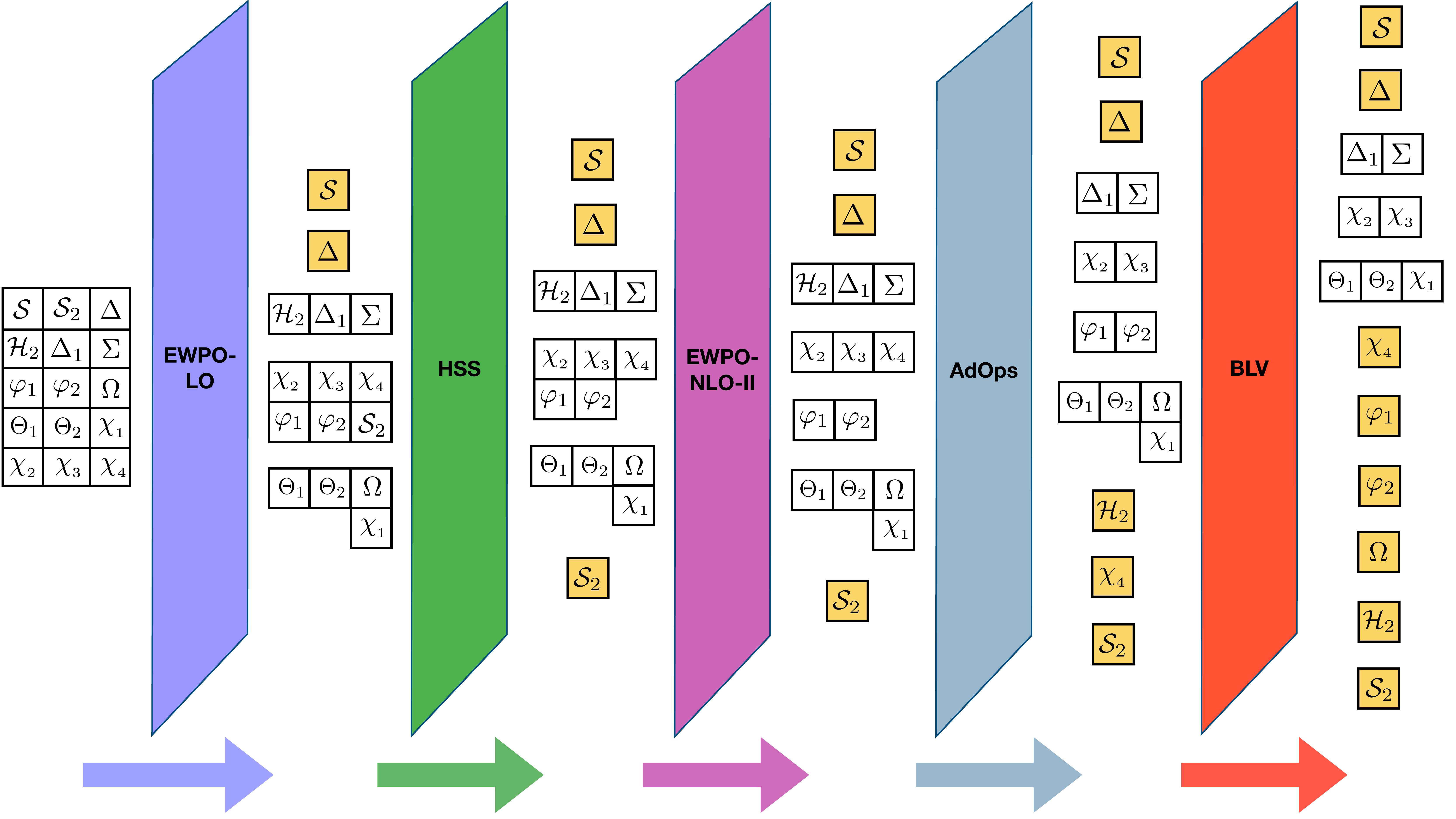}
	\caption{Cumulative Evolutions of BSM classification in the light of effective operators affecting EWPO-LO, HSS, EWPO-NLO-II, and also in presence of relevant Additional Operators (AdOps), and B,L violating (BLV) ones.}
	\label{fig:ops-class-curtain}
\end{figure}

Based on our discussion so far, we have observed that incorporating a single ``observable'' at a time leads to
a specific pattern for the BSM models, see Figs.~\ref{fig:degeneracy-breaking-EWPO-LO}, \ref{fig:degeneracy-breaking-EWPO-NLO-I}, \ref{fig:degeneracy-breaking-Higgs-data}, \ref{fig:BSMs-op-obs-EWPO-NLO-II},  \ref{fig:degeneracy-breaking-Add-Ops} for EWPO-LO, EWPO-NLO-I, EWPO-NLO-II, HSS, and AdOps respectively.  It is quite evident from these figures that the  different models are pooled together to form a degenerate class determined by a specific set of effective operators. 

Now the question that comes to mind is what is the result of this classification once we switch on all the ``observables''  simultaneously? This is a legitimate query if we assume that all observables probe the same energy scale. In Fig.~\ref{fig:ops-class-curtain} we show the combined effect of taking all observables into account. It is worth mentioning that the intermediate BSM classifications depends on the ordering at which the observables are considered, but the final result remains unaffected by that. 

The 15 BSM scenarios can be finally classified in relation to the observables, see Eqs.~\ref{eq:ObsEffOps-ewpolo}-\ref{eq:ObsEffOps-blv}, as
\begin{eqnarray}\label{eq:final-BSM-class}
\text{Class-F1:} & \Phi_i^{\text{F1}} \in \{ \Delta_{1}, \Sigma\} &\Rightarrow  \text{34 operators};\nonumber\\
\text{Class-F2:} &  \Phi_i^{\text{F2}} \in \{\Theta_1, \Theta_2,\chi_{_1}\} &\Rightarrow  \text{35 operators};\nonumber\\
\text{Class-F3:} &  \Phi_i^{\text{F3}} \in \{\chi_{_2},\chi_{_3}\} &\Rightarrow \text{15 operators};\\
\text{Class-F4:} &  \{\mathcal{S}_2\} &\Rightarrow \text{13 operators};\nonumber\\
\text{Class-F5:} &   \{\Delta\} &\Rightarrow  \text{14 operators};\nonumber\\
\text{Class-F6:} &   \{\mathcal{S}\} &\Rightarrow  \text{9 operators};\nonumber\\
\text{Class-F7:} &  \{\phi_1\} &\Rightarrow \text{35 operators};\nonumber\\
\text{Class-F8:} &   \{\phi_2\} &\Rightarrow  \text{32 operators};\nonumber\\
\text{Class-F9:} &   \{\Omega\} &\Rightarrow  \text{36 operator};\nonumber\\
\text{Class-F10:} &   \{\chi_{_4}\} &\Rightarrow  \text{32 operator};\nonumber\\
\text{Class-F11:} &   \{\mathcal{H}_2\} &\Rightarrow  \text{37 operators}\nonumber.
\end{eqnarray}

In summary, we have sketched the interplay of sets of dimension-6 operators with precision observables and how they emerged from each of the 15 BSM scenarios up to 1-loop. Based on this interplay, we have classified these new physics scenarios and pooled together into 11 independent classes (F1-F11),  see Eq.~\ref{eq:final-BSM-class}. Within each of these classes, the models impact the observables in the same way. Each of these classes can be clearly discriminated from one another, i.e. they each possess a unique signature of operators. However, some of the classes consist of multiple models, e.g. F1, F2, and F3, for which clear discrimination based on how their operators affect observables cannot be devised. Thus, the computation of dimension-6 operators up to 1-loop and the observables considered here are not sufficient to break the degeneracy between these models. To do so, one needs to go beyond 1-loop, and one needs to incorporate operators beyond dimension-6. After breaking the degeneracy in the model space entirely, i.e. by distinguishing each of the BSMs through a unique signature of operators that can be probed experimentally, it will be possible to address the so-called {\it inverse problem} of collider phenomenology.

\section{Conclusions}\label{sec:conclusion}

In recent years, it has become increasingly popular to tension experimental measurements with theory predictions using the language of Effective Field Theories (EFTs). This well-established theory framework aims to remain as model-independent as possible when giving measurements an interpretation in terms of new physics. However, the complexity of the operator space, already at dimension-6 within SMEFT, and the limited precision with which operators can be constrained experimentally jeopardise the applicability of this framework if no assumptions on the Standard Model extensions are imposed. Instead, we argue that a combined effort that links the bottom-up EFT approach with top-down UV model-building ideas will significantly improve our ability to learn from data gathered at collider experiments. 
In this paper, we have considered 15 different single scalar field extensions of the SM as example UV scenarios. These include real and complex colored and uncolored particles, and thus encapsulate most of the phenomenologically relevant scalar-extended new physics interactions. We have implemented all of these scenarios in CoDEx \cite{Bakshi:2018ics} and integrated out the heavy BSM particles up to 1-loop level, thereby generating their full sets of dimension-6 effective operators. To obtain an unambiguous operator basis we have employed the equation of motions and have taken into account light-heavy loop-propagators, in addition to the usual heavy loop-propagators, to compute the effective operators. We have tabulated the dominant contributions of all the effective operators for each such BSM model. To adjudge the impact of these operators, we have investigated whether they impact precision observables, like EWPO and Higgs Signal into four categories: (i) EWPO-LO, (ii) EWPO-NLO-I, (iii) Higgs Signal Strength, (iv) EWPO-NLO-II. 
In addition to these observables, we consider $B, L$ conserving (AdOps) and violating (BLV) operators that emerge in the process of integration out the heavy degrees of freedom. Considering all of them we define our set of observables. First, we have shown how these 15 models can be pooled into independent classes for each of the observables separately. We have further discussed the interrelation between different classes for each of the observables. Then, we have applied all the observables simultaneously to find the final classification of BSM scenarios. In the end, one is still left with a small number of classes that cannot be discriminated using the observables considered. Here, we have restricted ourselves to a complete 1-loop computation of the effective dimension-6 operators. One way forward to achieving a full separation of all models would require to consider higher-dimensional operators or include further observables. Thus, this approach of classifying UV models provides guidance on where to truncate the perturbative series while calculating the set of effective operators that are tensioned against experimental measurements. The outlined methodology can be straightforwardly extended to scenarios without degenerate mass spectrum in the UV \cite{Anisha:2019nzx, Banerjee:2020bym,Banerjee:2020jun}.

\section{Acknowledgement}
	The work of S.D.B., J.C. is supported by the Science and Engineering Research Board,
Government of India, under the agreements SERB/PHY/2016348  and SERB/PHY/2019501 and Initiation Research Grant, agreement
number IITK/PHY/2015077, by IIT Kanpur. M.S. is supported by the STFC under grant
ST/P001246/1. We thank the IPPP for access to its computing cluster.


	\providecommand{\href}[2]{#2}
	\addcontentsline*{toc}{section}{}
	\bibliographystyle{JHEP}
	\bibliography{refs-EFT}

	
\end{document}